\documentclass[11pt]{article}

\usepackage[margin=1in]{geometry}

\usepackage{setspace}
\usepackage{titlesec}

\renewcommand\thesection{\Roman{section}}
\renewcommand\thesubsection{\Alph{subsection}}
\renewcommand\thesubsubsection{\arabic{subsubsection}}
\renewcommand\theparagraph{\alph{paragraph}}

\titleformat{\section}
  {\normalfont\bfseries}
  {\thesection.}
  {0.75em}
  {\MakeUppercase}

\titleformat{\subsection}
  {\normalfont\bfseries}
  {\thesubsection.}
  {0.75em}
  {}

\titleformat{\subsubsection}
  {\normalfont\bfseries}
  {\thesubsubsection.}
  {0.75em}
  {}

\titleformat{\paragraph}
  {\normalfont\bfseries}
  {\theparagraph.}
  {0.75em}
  {}

\titlespacing*{\paragraph}{0pt}{1.25ex plus .2ex}{1em}

\usepackage{graphicx}
\usepackage{amsmath,amssymb}
\usepackage{physics}
\usepackage{bm}
\usepackage[numbers]{natbib}
\usepackage{hyperref}

\title{Achieving Extraordinary Acoustic Transmission in a Single Slit by Boundary Impedance Engineering}
\author{J. Sumaya-Martinez and J. Mulia-Rodriguez\\Facultad de Ciencias, UAEMex, Toluca, Mexico}
\date{}

\begin{document}
\maketitle

\begin{abstract}
Extraordinary acoustic transmission (EAT) through subwavelength apertures is typically achieved using periodic arrays or external resonant elements. Here we demonstrate that strong and tunable EAT can be realized in a single isolated subwavelength slit by introducing one or more internal geometric constrictions. These constrictions act as acoustic impedance transformers that simultaneously modify the effective acoustic length of the slit and its coupling to the surrounding medium. Using a one-dimensional transmission-line model supported by numerical simulations, we show that near-unity transmission can be achieved at deeply subwavelength wavelengths in ultra-thin rigid plates. Multi-constriction geometries further give rise to resonance splitting and multi-peak transmission spectra, revealing a universal impedance-based mechanism shared by acoustic and electromagnetic extraordinary transmission phenomena.
\end{abstract}

\section{Introduction}

Acoustic transmission through apertures whose lateral dimensions are much smaller than the wavelength is fundamentally limited by strong impedance mismatch between the radiating field in the surrounding fluid and the guided mode supported inside the aperture. As a result, a single subwavelength slit or hole in a rigid plate typically exhibits weak transmission, even when the aperture supports a cutoff-free plane mode. Over the past two decades, however, it has been shown that this limitation can be overcome through carefully engineered geometries that give rise to \emph{extraordinary acoustic transmission} (EAT), in close analogy with the phenomenon of extraordinary optical transmission (EOT) discovered in perforated metallic screens \citep{Ebbesen1998}.

In acoustics, several mechanisms enabling EAT have been identified. Periodic arrays of subwavelength apertures can support collective resonances and diffraction-mediated effects that strongly enhance transmission at selected frequencies \citep{Lu2007,EstradaWaveMotion2011}. In related configurations, the coupling of aperture modes to surface-wave-like fields can lead to enhanced transmission and beaming \citep{Christensen2010PRB}. A complementary route relies on embedding resonant substructures—such as Helmholtz resonators or labyrinthine/coiled channels—within or around the apertures to tailor the effective impedance and phase response, enabling high transmission in ultra-thin structures \citep{Koju2014,Liang2013}. More recently, giant extraordinary transmission has been demonstrated in engineered platforms where Fabry--P\'erot-like resonances and impedance matching are jointly controlled \citep{Devaux2020}.

A unifying interpretation of these observations is that extraordinary transmission is, at its core, an \emph{impedance-matching} problem: strong transmission occurs when the input impedance of the structured aperture matches the radiation impedance of the surrounding medium. In a uniform subwavelength slit this condition is rarely met, as the characteristic impedance of the guided mode inside the slit differs strongly from that of free space. Most existing strategies therefore engineer impedance matching indirectly, either through collective effects requiring periodicity or through external resonant elements that increase geometrical complexity.

Here we show that strong and tunable EAT can be achieved in a \emph{single isolated} subwavelength slit by engineering the impedance \emph{inside} the slit itself. Specifically, we introduce one or more narrow internal constrictions (``ridges'') that act as \emph{acoustic impedance transformers}. These internal discontinuities simultaneously (i) modify the effective acoustic length of the slit cavity and (ii) transform the input impedance seen at the slit entrance, enabling near-unity transmission at deeply subwavelength wavelengths without periodicity or external resonators. Using a one-dimensional transmission-line model supported by numerical simulations, we demonstrate single-peak EAT and, for multiple constrictions, coupled-cavity resonance splitting and multi-peak spectra.

Beyond providing a compact route to EAT in ultra-thin plates, the mechanism presented here highlights the universal role of impedance transformation in extraordinary transmission phenomena across acoustic and electromagnetic wave systems.

\section{Theoretical model}

We consider a normally incident plane acoustic wave interacting with a rigid plate of thickness $h$ perforated by a single subwavelength slit that is invariant in the transverse direction. For wavelengths much larger than the slit width, only the fundamental plane mode propagates inside the slit and the problem can be modeled as a one-dimensional acoustic transmission line \citep{Kinsler2000}. We neglect viscothermal losses to isolate the underlying mechanism; their main effect is to broaden resonances and reduce peak transmission, but they do not change the impedance-transformation picture discussed below.

\subsection{Uniform slit as a reference}
For a uniform slit of cross-sectional area $S_1$, the guided mode has wavenumber
\begin{equation}
k = \omega/c,
\end{equation}
and characteristic impedance
\begin{equation}
Z_1 = \rho c / S_1,
\end{equation}
where $\rho$ and $c$ denote density and sound speed of the host fluid. In a simple lossless model, the input impedance of a uniform slit section of length $h$ terminated by a radiation impedance can be written in terms of standard transmission-line relations. In practice, strong mismatch between the slit impedance and the external radiation impedance leads to weak coupling and low transmission except near Fabry--P\'erot (FP) conditions \citep{Christensen2008PRL}.

\subsection{Single internal constriction}
We introduce an internal constriction of length $d$ and area $S_2<S_1$, with impedance
\begin{equation}
Z_2 = \rho c / S_2 > Z_1.
\end{equation}
The structured slit is described by cascading transfer matrices for each uniform segment,
\begin{equation}
\mathbf{T}_j =
\begin{pmatrix}
\cos(k\ell_j) & i Z_j \sin(k\ell_j)\\
i Z_j^{-1}\sin(k\ell_j) & \cos(k\ell_j)
\end{pmatrix},
\end{equation}
with $(\ell_j,Z_j)$ the length and impedance of the $j$th segment. From the total transfer matrix one obtains the input impedance $Z_{\mathrm{in}}$ and the transmitted power coefficient.

Although the full expression is cumbersome, the subwavelength regime yields a transparent physical picture: the constriction acts as an impedance transformer and introduces an effective length correction. To leading order, the FP condition is shifted to
\begin{equation}
k\left(h+\Delta h_{\mathrm{eff}}\right)\approx n\pi,
\end{equation}
with an effective length increment scaling as
\begin{equation}
\Delta h_{\mathrm{eff}} \sim d\left(\frac{Z_2}{Z_1}-1\right),
\end{equation}
which accounts for the redshift observed in the numerical spectra. Simultaneously, the impedance discontinuity transforms the entrance impedance; for a single stepped-impedance section a compact approximation is
\begin{equation}
Z_{\mathrm{in}} \approx Z_1 \frac{Z_2 + i Z_1 \tan(k d)}{Z_1 + i Z_2 \tan(k d)}.
\end{equation}
When $Z_{\mathrm{in}}$ approaches the radiation impedance of the surrounding medium at resonance, strong transmission becomes possible.

\subsection{Multiple constrictions}
The formalism extends directly to multiple constrictions, yielding cascaded impedance transformations and coupled-cavity behavior. As shown in Sec.~\ref{sec:results}, two constrictions separated by a uniform segment produce resonance splitting characteristic of hybridized cavity modes.

\section{Numerical results}

\label{sec:results}
We summarize representative parametric sweeps obtained with the one-dimensional transmission-line model for a rigid slit containing one or more internal constrictions. The plate thickness $h$ is used as the reference length and wavelengths are normalized as $\lambda/h$.

\subsection{Single internal constriction: ridge length}
\begin{figure}[!ht][t]
\centering
\includegraphics[width=0.9\linewidth]{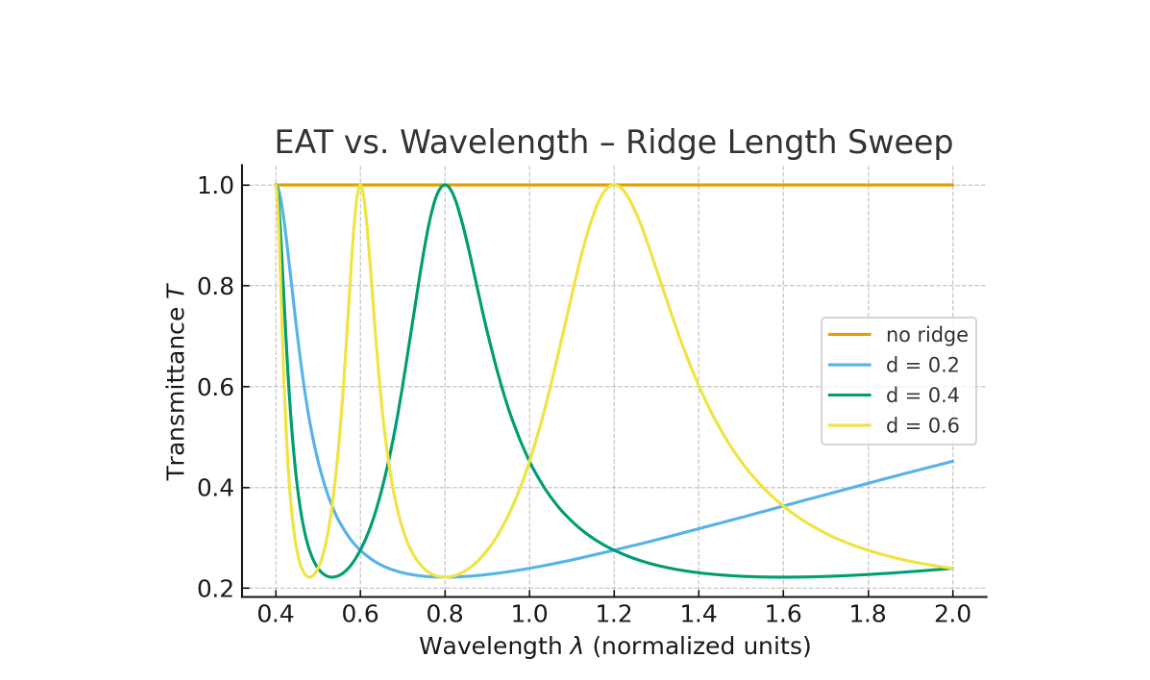}
\caption{Power transmittance $T$ as a function of normalized wavelength $\lambda/h$ for a single subwavelength slit containing one internal constriction of length $d$. Curves correspond to several values of $d/h$ at fixed impedance contrast $Z_2/Z_1=4$. The uniform slit ($d=0$) exhibits weak transmission, while a finite constriction yields a strong extraordinary-transmission peak that redshifts as $d/h$ increases due to an effective increase of the acoustic cavity length.}
\label{fig:fig1}
\end{figure}

Figure~\ref{fig:fig1} shows the power transmittance $T$ as a function of normalized wavelength for several ridge lengths $d/h$ at fixed impedance contrast $Z_2/Z_1=4$. The uniform-slit case ($d=0$) yields weak and nearly flat transmission, consistent with strong impedance mismatch. Introducing a finite constriction produces a pronounced extraordinary-transmission peak that redshifts with increasing $d/h$, in agreement with the effective length correction discussed in Sec.~II. For intermediate ridge lengths the peak transmission approaches unity, indicating efficient impedance matching.

\subsection{Single internal constriction: impedance contrast}
\begin{figure}[!ht][t]
\centering
\includegraphics[width=0.9\linewidth]{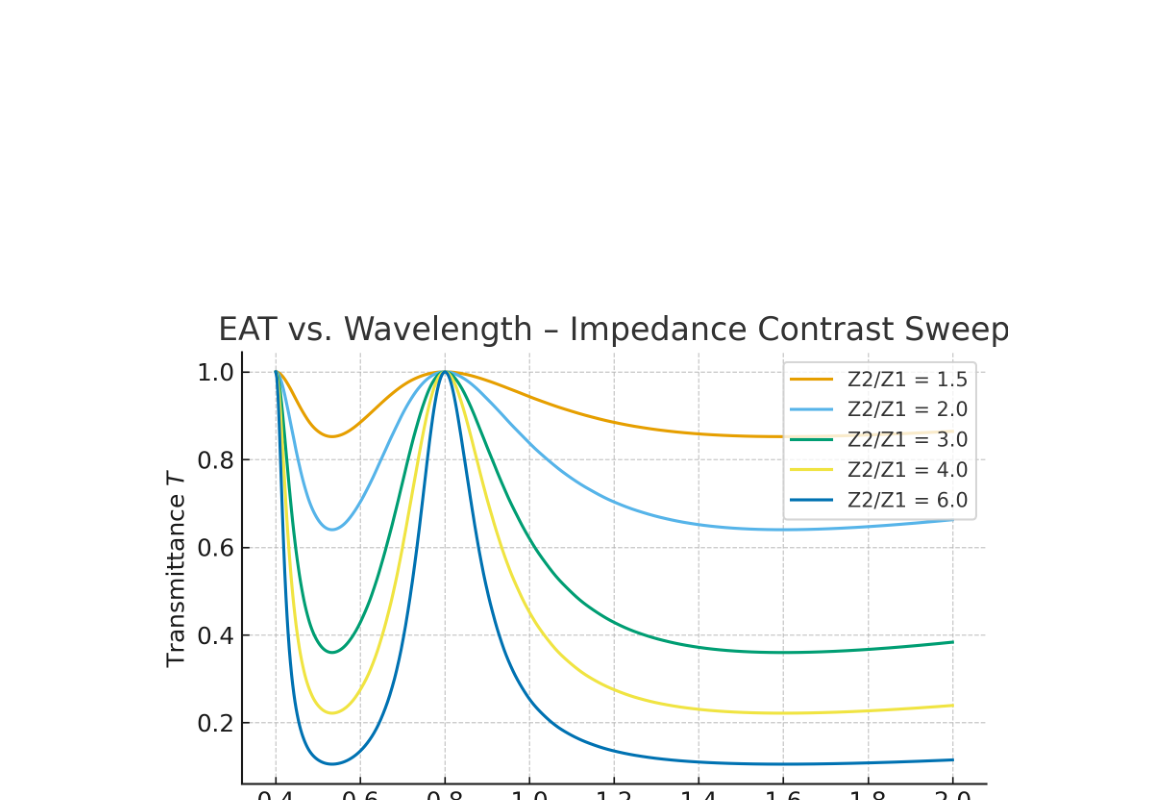}
\caption{Power transmittance $T$ as a function of normalized wavelength $\lambda/h$ for a single internal constriction of fixed length $d/h=0.4$, and different impedance contrasts $Z_2/Z_1$. Moderate contrasts yield sharp, high-transmission resonances, while very large contrasts reduce the peak transmission due to stronger reflection at the impedance discontinuities.}
\label{fig:fig2}
\end{figure}

Figure~\ref{fig:fig2} shows the effect of varying the impedance contrast $Z_2/Z_1$ at fixed ridge length $d/h=0.4$. Moderate contrasts produce sharp, high-transmission peaks, whereas very large contrasts reduce peak transmission due to increased reflection at the impedance discontinuities. The resonance position also shifts toward longer wavelengths as $Z_2/Z_1$ increases, consistent with increased phase accumulation inside the high-impedance segment.

\subsection{Two internal constrictions: separation and mode splitting}
\begin{figure}[!ht][t]
\centering
\includegraphics[width=0.9\linewidth]{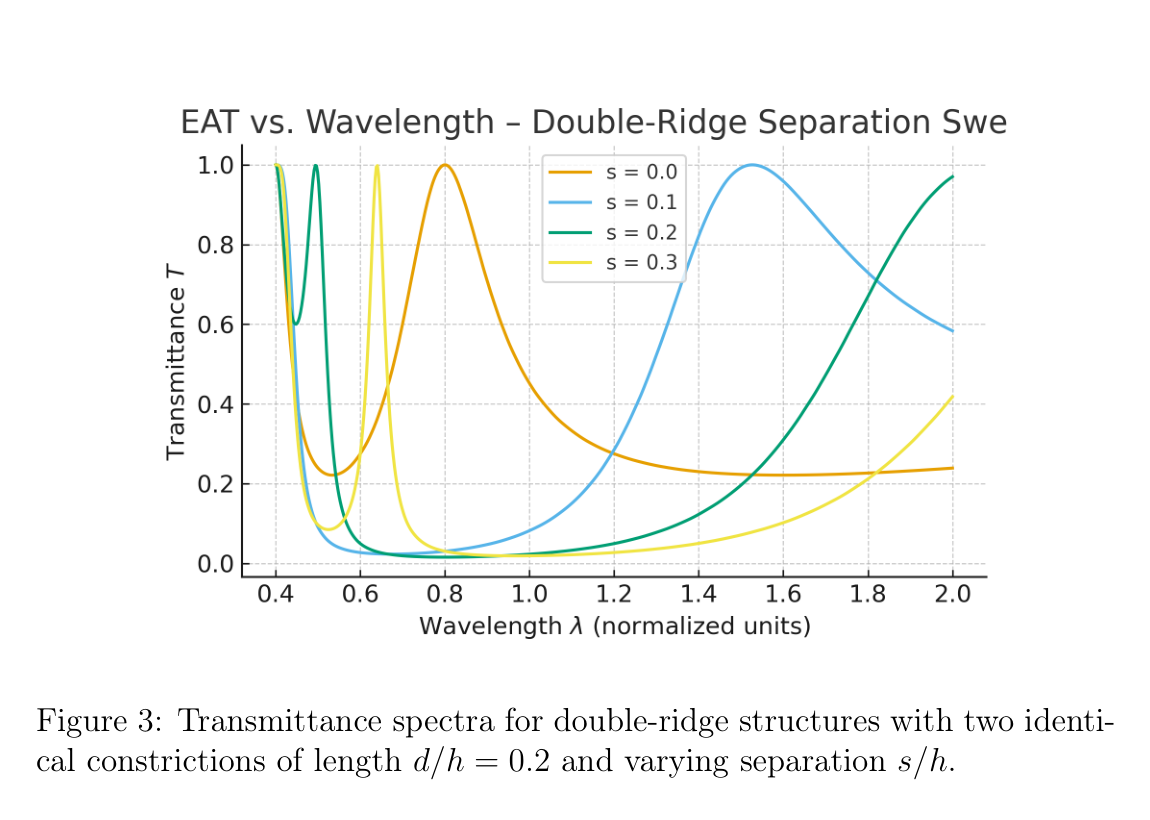}
\caption{Power transmittance $T$ as a function of normalized wavelength $\lambda/h$ for a subwavelength slit containing two identical internal constrictions of length $d/h=0.2$ separated by a distance $s$. For $s=0$ the system behaves as a single longer constriction and supports one extraordinary-transmission peak. Increasing $s$ leads to resonance splitting, characteristic of coupled-cavity behavior.}
\label{fig:fig3}
\end{figure}

Figure~\ref{fig:fig3} displays spectra for double-ridge structures composed of two identical constrictions of length $d/h=0.2$ separated by a central segment of length $s/h$. For zero separation the two ridges behave as a single longer constriction and produce a single EAT peak. As the separation increases, the resonance splits into two peaks, characteristic of coupled-cavity hybridization. Importantly, high transmission is retained, demonstrating that the impedance-matching mechanism persists in multi-resonant configurations.

\section{Discussion}

The results confirm that internal constrictions in a single subwavelength slit provide a robust and tunable route to extraordinary acoustic transmission via distributed impedance transformation and effective length corrections. The one-dimensional model is valid when the transverse dimensions remain deeply subwavelength so that higher-order modes are evanescent. In practice, viscothermal losses will reduce peak transmission for very narrow constrictions, but the predicted trends---resonance redshift with $d$ or $Z_2/Z_1$, and resonance splitting in multi-constriction geometries---are expected to persist.

Compared with array-based EAT \citep{Lu2007,EstradaWaveMotion2011}, the present mechanism is intrinsically local and does not require periodicity or surface-wave coupling. Compared with Helmholtz- or coiling-based ultrathin designs \citep{Koju2014,Liang2013}, the resonances retain a Fabry--P\'erot character while achieving strong impedance matching through a stepped-impedance transformer inside the slit. This impedance viewpoint also clarifies the close analogy with EOT in metallic screens \citep{Ebbesen1998,GarciaVidal2001PRL}, where enhanced transmission can similarly be interpreted through resonant tunneling and impedance matching in structured apertures.

From a design perspective, internal impedance structuring enables compact narrowband or multiband acoustic transmission devices in ultra-thin rigid plates. By controlling the number, lengths, and separations of constrictions, one can engineer transmission spectra ranging from single sharp peaks to split resonances suitable for multi-frequency filtering and sensing.

\section{Conclusions}

We have demonstrated that extraordinary acoustic transmission can be achieved in a single isolated subwavelength slit by introducing internal geometric constrictions that act as acoustic impedance transformers. A transmission-line framework shows that the constrictions simultaneously (i) increase the effective acoustic length of the slit cavity and (ii) transform the entrance impedance toward the radiation impedance of the surrounding medium, enabling near-unity transmission at deeply subwavelength wavelengths in ultra-thin rigid plates. Multiple constrictions naturally produce coupled-cavity mode splitting and multi-peak extraordinary-transmission spectra. The results highlight a universal impedance-based mechanism that connects acoustic extraordinary transmission with electromagnetic enhanced transmission in structured apertures, and provide a simple route to compact acoustic filters, sensors, and impedance-matched interfaces.

\section*{Author Declarations}

\subsection*{Conflict of Interest}
The authors declare that there is no conflict of interest regarding the publication of this article.

\subsection*{Data Availability}
The data that support the findings of this study are available from the corresponding author upon reasonable request.

\bibliographystyle{plainnat}
\bibliography{references}

\end{document}